# Electrical transport properties of $0.5Li_2O$-$0.5M_2O$-$2B_2O_3$ (M=Li, Na and K) glasses

**G. Paramesh, Rahul Vaish and K. B. R. Varma***

Materials Research Centre,

Indian Institute of Science,

Bangalore 560 012,

India.

---

*Corresponding Author; E-Mail: kbrvarma@mrc.iisc.ernet.in;

FAX: 91-80-23600683; Tel. No: 91-80-22932914

**Abstract:**

Transparent glasses in the system $0.5Li_2O$-$0.5M_2O$-$2B_2O_3$ (M=Li, Na and K) were fabricated via the conventional melt quenching technique. Amorphous and glassy nature of the samples were confirmed via the X-ray powder diffraction and the differential scanning calorimetry, respectively. The frequency and temperature dependent characteristics of the dielectric relaxation and the electrical conductivity were investigated in the 100 Hz - 10 MHz frequency range. The imaginary part of the electric modulus spectra was modeled using an approximate solution of Kohrausch-Williams-Watts relation. The stretching exponent, $\beta$, was found to be temperature independent for $0.5Li_2O$-$0.5Na_2O$-$2B_2O_3$ (LNBO) glasses. The activation energy associated with DC conductivity was found to be higher (1.25eV) for $0.5Li_2O$-$0.5K_2O$-$2B_2O_3$ (LKBO) glasses than that of the other glass systems under study. This could be attributed to the mixed cationic effect.

## 1. Introduction

Noncentrosymmetric borate based compounds are found to be promising for piezoelectric, pyroelectric and non-linear optical based device application. Various borate based single crystals like $BaB_2O_4$ [1], $BiB_3O_6$ [2], $CsLiB_6O_{10}$ [3] have been investigated and reported to be promising from their physical properties point of view. Borate based compounds have attained importance mainly because of their nonlinear optical properties in generating visible and UV light apart from their high laser damage threshold, wide optical transmission window and low cost involved in material processing. $Li_2B_4O_7$ [4], $LiNaB_4O_7$ [5] and $LiKB_4O_7$ [6] have been reported to be promising for non-linear optical applications. $Li_2B_4O_7$ crystal belongs to the space group $I4_1cd$ whereas the other two crystallize in orthorhombic space groups ($LiNaB_4O_7$ in Fdd2 and $LiKB_4O_7$ in $P2_12_12_1$ space groups). Though the single crystals of these are superior in their physical properties, the melt growth process is tedious time consuming and expensive. Therefore, it is necessary to look for alternatives to the single crystals and see whether these exhibit physical properties that are akin to that of single crystals. One of the alternatives would be to going glass ceramic route of fabricating transparent materials. Glasses of the same materials are being investigated to be used in optical devices, especially in the UV range. There are many advantages in the use of glasses as the process of fabrication involved is easy and less expensive than that of single crystals. The transparent glass nano/micro crystal composites have attracted the many researchers for demonstration of symmetry dependent properties like pyroelectricity, piezoelectricity and second harmonic generation [7-9]. The controlled heat treatment of glasses is the way to obtain the glass nano/micro crystal composites of polar crystals. In the process of fabricating transparent glass nano/micro crystal composites of $Li_2B_4O_7$, $LiNaB_4O_7$ and $LiKB_4O_7$ for their multifunctionalities which include piezoelectric, pyroelectric and nonlinear optical properties; to begin with transparent glasses of $0.5Li_2O$-$0.5M_2O$-$2B_2O_3$ (M= Li, Na and K) were made and the comparative study of the electrical transport properties over wide range of temperatures and frequencies was carried out. Further the compositions in the above system were chosen such that one eventually obtains $Li_2B_4O_7$, $LiNaB_4O_7$ and $LiKB_4O_7$ crystalline phases on crystallization.

**2. Experimental**

Transparent glasses in the system $0.5Li_2O$-$0.5M_2O$-$2B_2O_3$ (M=Li, Na and K) were fabricated via the conventional melt quenching technique. For this $Li_2CO_3$ (99.9 % Merck), $Na_2CO_3$ (99.9 % Merck), $K_2CO_3$ (99.9 % Merck) and $H_3BO_3$ (99.9% Merck) were used. Each composition was mixed and melted in a platinum crucible at 1000$^oC$ for 30 min. Melts were quenched by pouring on a steel plate and pressed with another plate to obtain 1-1.5 mm thick glass plates. X-ray powder diffraction study was performed at room temperature on the as-quenched samples to confirm their amorphous nature. The glassy nature of the as-quenched samples was confirmed by subjecting the samples to non-isothermal Differential Scanning Calorimetric (Perkin Elmer, Diamond DSC) studies, in the 350-650$^oC$ temperature range.

The as-quenched glass samples were painted with silver and silver epoxy was used to bond silver leads to the samples for the capacitance and dielectric loss (D) measurements. The capacitance (C) and dielectric loss ( $D = \varepsilon_r^{''} / \varepsilon_r^{'}$ ) were monitored as functions of both frequency (100 Hz-10 MHz) and temperature (50$^{o}$C-550$^{o}$C), using a HP4194A impedance/gain phase analyzer at a signal strength of 0.5 $V_{rms}$. Based on these data the real and imaginary parts of the dielectric constant were calculated by taking the dimensions and electrode geometry of the sample into account.

**3. Results and discussion**

X-ray powder diffraction patterns were recorded for pulverized as-quenched glass samples which confirm their amorphous nature (Fig.1). These as-quenched samples were transparent and colorless. The DSC traces that are obtained for $Li_2O$-$2B_2O_3$ (LBO), LNBO and LKBO glasses at a heating rate of 10$^{o}$C/min are shown in Fig. 2 (a-c). The DSC traces for the glasses exhibit endotherms followed by exotherms which are associated with the glass transition and the crystallization temperatures, respectively. The glass transition ($T_g$) and the onset of the crystallization ($T_{cr}$) temperatures are identified as the temperatures corresponding to the intersection of two linear portions of the transition elbows in the DSC traces. An exotherm in Fig. 2 (a) which corresponds to $Li_2O$-$2B_2O_3$ crystallization is sharper than that of the other glasses (LNBO and LKBO) under study. Generally the incidence of a sharp exothermic peak suggests the crystallization process to be bulk [10]. The glass transition and the crystallization temperatures for LBO, LNBO and LKBO glasses are reported in Table I. The glass transition temperature was found to be lower for LKBO glasses than that of the other two glass systems (LBO and LNBO).

This is because of more open structure (more non-bridging oxygen) associated with LKBO glasses due to presence of $K^+$ ions which have higher ionic radii than that of Na+ and Li+ ions. It results in the expansion of the network forming units due to the large voids which weaken the glass network in the mixed LKBO glasses as compared to that of the LNBO and LBO glasses. It is also noticed that the crystallization temperature was higher for LKBO glasses than that of the other two systems (viz LKBO>LNBO>LBO) in the present investigation. This is due to the fact that the $Li^+$, $K^+$ and $Na^+$ ions are randomly mixed in the respective glass matrices in all the diffusion pathways and each cation creates its own chemical environment [11]. These cation species prefer to migrate via pathways of sites adjusted to their own requirements. The ionic mobility is decreased if the pathways of the respective species interfere with each other. This blocking considerably reduces the ionic diffusion in LKBO and LNBO glasses (due to mixed alkali effect) in comparison to the corresponding single cation glasses (LBO). The large ionic radii mismatch between $Li^+$ and $K^+$ ions than $Li^+$ and $Na^+$ ions strengthen more ionic blocking in LKBO glasses than LNBO glasses hence requires higher thermal activation for crystallization than the other glass system under study. This also facilitates surface crystallization (interface controlled) as compared to that of the bulk crystallization (diffusion controlled) in LKBO glasses which is evident from the broad exotherm associated with these glasses (Fig. 2 (c)). It indicates that LKBO glasses have more pronounced mixed alkali effect than LNBO glasses. The glass transition and the crystallization temperatures are usually employed to estimate the thermal stability. The relative location of $T_{cr}$ with reference to $T_g$ in the DSC/DTA trace is a measure of the thermal stability of glasses. The thermal stability ($T_{cr}$-$T_g$) of the glasses is a crucial parameter to be noted from their technological applications point of view. Glasses should be sufficiently stable against crystallization. However, one must be able to form nuclei and subsequently grow crystals within the glass matrix on heat treatment inorder to obtain glass-ceramics. The higher values of the $T_{cr}$-$T_g$ delay the nucleation process. In the present investigations, LKBO glasses have higher thermal stability ($T_{cr}$-$T_g$) than the other glass systems under study.

The variation of the dielectric constant ($\varepsilon_r^{'}$) with frequency (100 Hz-10 MHz) of measurement for LBO glasses at different temperatures is shown in Fig. 3 (a). At all the temperatures under study, $\varepsilon_r^{'}$ decreases with increase in frequency. The decrease is significant especially at low frequencies which may be associated with the mobile ion polarization combined with electrode polarization. The low frequency dispersion of $\varepsilon_r^{'}$ gradually increases with increase in temperature which may be due to an increase in the electrode polarization as well as thermal activation of $Li^+$ ions in the LBO glasses. As the frequency increases, $\varepsilon_r^{'}$ decreases due to high periodic reversal of the field, which reduces the contribution of the charge carriers towards the dielectric constant. Similar trends were found for the other two glasses (LNBO and LKBO) which are not shown in the figure.

The variation of the dielectric loss (*D*) with the frequency at various temperatures is shown in Fig. 3 (b) (with silver paint electrodes) for LBO glasses. It increases with increase in temperature, which is attributed to the increase in electrical conductivity of the glasses. Interestingly, relaxation peaks were encountered when the measurements were done at 225$^o$C and 250$^o$C. These relaxation peaks are due to electrode related polarization and electrode material dependent [12]. Whereas, in the dielectric loss spectra no relaxation peaks were observed for LNBO and LKBO glasses (not depicted in the Fig. 3(b)). In order to compare dielectric behavior of these glasses (LBO, LNBO and LKBO), dielectric constant and the loss spectra at 250$^{o}$C are shown in the Figs. 4 (a) and 4 (b), respectively. The dielectric dispersion is found to be less for LKBO glasses than that of LNBO and LBO glasses. This is because of mixed alkali effect associated with these glasses which delayed the ionic ($Li^+$ and $K^+$) response to the external electric field. The loss decreases with increase in frequency (Fig.4 (a)) for all the glasses under study. It is also observed that the dielectric loss is more for LBO glasses as compared to that of LNBO and LKBO glasses. It is attributed to the higher conductivity in LBO glasses and consistent with that of the dielectric constant behavior.

In order to have further insight into the dielectric behavior of these glasses, the electric modulus formalism was invoked. The use of electric modulus approach helps in understanding the bulk response of materials. Therefore, this approach could effectively be employed to separate out electrode effects. This would also facilitate to circumvent the problems caused by electrical conduction which might mask the dielectric relaxation processes. The complex electric modulus ($M^*$) is defined in terms of the complex dielectric constant ($\varepsilon$*) and is represented as [13]:

$$M^* = (\varepsilon^*)^{-1} \tag{1}$$

$$M' + iM'' = \frac{\varepsilon_r^{'}}{(\varepsilon_r^{'})^2 + (\varepsilon_r^{''})^2} + i\frac{\varepsilon_r^{''}}{(\varepsilon_r^{'})^2 + (\varepsilon_r^{''})^2} \tag{2}$$

where $M^{'}$, $M^{''}$ and , $\varepsilon_r^{'}$, $\varepsilon_r^{''}$ are the real and imaginary parts of the electric modulus and dielectric constants, respectively. The real and imaginary parts of the modulus at different temperatures are calculated using Eq. 2 for the LBO glasses under study and are depicted in Figs. 5 (a) and 5 (b), respectively. One would conclude from Fig.5 (a) that at low frequencies, $M'$ approaches zero at all the temperatures under study suggesting the suppression of the electrode polarization. $M'$ reaches a maximum value corresponding to $M_\infty$ = $(\varepsilon_\infty)^{-1}$ due to the relaxation process. The imaginary part of the electric modulus (Fig. 5 (b)) is indicative of the energy loss under electric field. The $M''$ peak shifts to higher frequencies with increasing temperature. This evidently suggests the involvement of temperature dependent relaxation processes in the LBO glasses. Similar trends were observed in LNBO and LKBO glasses which are not shown in the Fig. 5. The frequency regime that is below the $M''$ peak position indicates the range in which the ions drift to long distances. In the frequency range which is above that of the peak, the ions are spatially confined to potential wells and free to move within the wells. The frequency range where the peak occurs is suggestive of the transition from long-range to short-range mobility. The electric modulus could be expressed as the Fourier transform of a relaxation function $\phi(t)$ [14]:

$$M^* = M_\infty \left[ 1 - \int_0^\infty \exp(-\omega t) \left( -\frac{d\phi}{dt} \right) dt \right] \tag{3}$$

where the function $\phi(t)$ is the time evolution of the electric field within the materials and is usually taken as the Kohlrausch-Williams-Watts (KWW) function [15]:

$$\phi(t) = \exp\left[ -\left( t/\tau_m \right)^\beta \right] \tag{4}$$

where $\tau_m$ is the conductivity relaxation time and the exponent $\beta$ (0 1] indicates the deviation from Debye type relaxation. The value of $\beta$ could be determined by fitting the experimental data in the above equations. But it is desirable to reduce the number of adjustable parameters while fitting the experimental data. Keeping this point in view, the electric modulus behavior of the present glass system is rationalized by invoking modified KWW function suggested by Bergman. The imaginary part of the electric modulus ($M''$) could be defined as [16]:

$$M'' = \frac{M''_{\max}}{(1-\beta) + \frac{\beta}{1+\beta}\left[\beta(\omega_{\max}/\omega) + (\omega/\omega_{\max})^{\beta}\right]} \quad (5)$$

where $M''_{\max}$ is the peak value of the $M''$ and $\omega_{max}$ is the corresponding frequency. The above equation (Eq. 5) could effectively be described for $\beta \geq 0.4$. Theoretical fits of Eq. 5 to the experimental data are illustrated in Fig. 5 (a) as the solid lines. The experimental data are well fitted to this model except in the high frequency regime. From the fitting of $M''$ versus frequency plots, the value of $\beta$ was determined. The value of $\beta$ is found to be temperature dependent for LBO glasses and shown in the Fig. 6. The fitted values of β for LNBO and LKBO glasses at various temperatures are also depicted in the same figure (Fig. 6). The lower value of β for LNBO and LKBO glasses than LBO glasses indicate strong interaction between dissimilar ions present in glasses. The value of β for LBO, LNBO and LKBO glasses at 250$^{o}$C is mentioned in the Table I.

The relaxation frequency associated with the process was determined from the plot of $M''$ versus frequency. The activation energy involved in the relaxation process of ions could be obtained from the temperature dependent relaxation time as:

$$f_{\max} = f_o \exp\left(-\frac{E_R}{kT}\right) \quad (6)$$

where $E_R$ is the activation energy associated with the relaxation process, $f_o$ is the pre-exponential factor, $k$ is the Boltzmann constant and $T$ is the absolute temperature. Fig. 6 shows a plot between ln ($f_{max}$) and 1000/$T$ along (for LBO glasses) with the linear fit (solid line) to the above equation (Eq. 6). The value that is obtained for $E_R$ is 0.66 eV, which is ascribed to the motion of $Li^+$ ions and is consistent with one reported in the literature [17]. The value of activation energy associated with relaxation process for LNBO and LKBO glasses is shown in Table I. The significantly higher activation energy for LNBO and LKBO glasses could be explained using dynamic structure model (DSM) [18]. This model explained the competition between dissimilar mobile ions in their attempts to establish their conduction pathways. Simulation studies showed that this competition led to the fragmentation of pathways, and hence a sharp decrease in ionic diffusion.

It is of interest to investigate into the transport mechanism in present glasses. Therefore, the DC conductivity at different temperatures ($\sigma_{DC(T)}$), was calculated from the electric modulus data. The DC conductivity could be extracted using the expression [19]:

$$\sigma_{DC}(T) = \frac{\varepsilon_o}{M_\infty(T) * \tau_m(T)} \left[ \frac{\beta}{\Gamma\left( 1/\beta \right)} \right] \qquad (7)$$

where $\varepsilon_o$ is the free space dielectric constant, $M_\infty$ ($T$) is the reciprocal of high frequency dielectric constant and $\tau_m$ ($T$) is the temperature dependent relaxation time. Fig. 7 shows the DC conductivity data obtained from the above expression (Eq. 7) at various temperatures for LBO glasses. The activation energy for the DC conductivity is calculated from the plot of ln ($\sigma_{DC}$) versus 1000/$T$ (Fig. 7) for LBO glasses. The plot is found to be linear and fitted using the following Arrhenius equation,

$$\sigma_{DC}(T) = B\exp\left(-E_{DC}/kT\right) \tag{8}$$

where $B$ is the pre-exponential factor, $E_{DC}$ is the activation energy for the DC conduction. The activation energy is calculated from the slope of the fitted line and is found to be 0.77 eV for LBO glasses. The activation energies associated with the DC conductivity for LNBO and LKBO glasses were calculated using Eqs. 7 and 8 and given in Table I. These values are in close agreement with that of corresponding activation energy associated with relaxation process. It suggests that similar energy barriers are involved in both the relaxation and conduction processes in all the glasses under study.

Figs. 8 (a) and 8 (b) show the real and imaginary parts of the electric modulus for LBO, LNBO and LKBO glasses at 250°C. Solid lines in Fig. 8 (b) are the theoretical fits to the Eq. 5. The relaxation frequency for all the glasses is shown in Table I. The relaxation was found to be at lower frequency for LKBO glass than that of LNBO and LBO glasses (viz. $f_{max}^{LKBO} < f_{max}^{LNBO} < f_{max}^{LBO}$). This is due to higher ionic mass of $K^+$ ions which can be relaxed at lower frequency than that of $Na^+$ and $Li^+$ ions.

AC conductivity at different frequencies was determined by using the dielectric data using the following formula:

$$\sigma_{AC} = \omega\varepsilon_o D\varepsilon_r' \tag{9}$$

where $\sigma_{AC}$ is the AC conductivity at a frequency $\omega$ (=$2\pi f$). Inorder to compare electrical conductivity of the LBO, LNBO and LKBO glasses, the frequency dependence of conductivity plots at 250°C for all three glasses are depicted in Fig. 9. It is clear from the figure that the conductivity is much lower for LKBO glasses than that of the LBO and LNBO glasses in the entire frequency range under study suggesting that in LKBO glasses the mixed alkali effect is stronger than that of LNBO glasses.

## 4. Conclusions

Electrical behavior of the transparent $0.5Li_2O$-$0.5M_2O$-$2B_2O_3$ (M=Li, Na and K) glasses were studied. The LBO glasses were found to be less stable thermally as compared to that of LKBO glasses. The electrical relaxation was rationalized using the electric modulus formalism. LKBO glasses were found to be electrically less conductive than that of LBO and LNBO glasses which may be of considerable interest from view point of electronic device applications.

Table 1: Physical characteristics for LBO, LNBO and LKBO glasses

| **Parameters** | **LBO** | **LNBO** | **LKBO** |
|---|---|---|---|
| $T_g$ ($^oC$) ($10^oC$/min) | 487 | 443 | 413 |
| $T_{cr}$ ($^oC$) ($10^oC$/min) | 535 | 545 | 567 |
| $f_{max}$ (Hz) ($250^oC$) | 3981072 | 31623 | 1122 |
| $\beta$ ($250^oC$) | 0.626 | 0.571 | 0.567 |
| $\sigma_{DC}$ ($\Omega^{-1}m^{-1}$)($250^oC$) | $3.3 \times 10^{-3}$ | $9.87 \times 10^{-6}$ | $3.44 \times 10^{-7}$ |
| $E_{DC}$ (eV) | 0.77 | 1.2 | 1.25 |
| $E_{relax}$ (eV) | 0.66 | 1.14 | 1.13 |

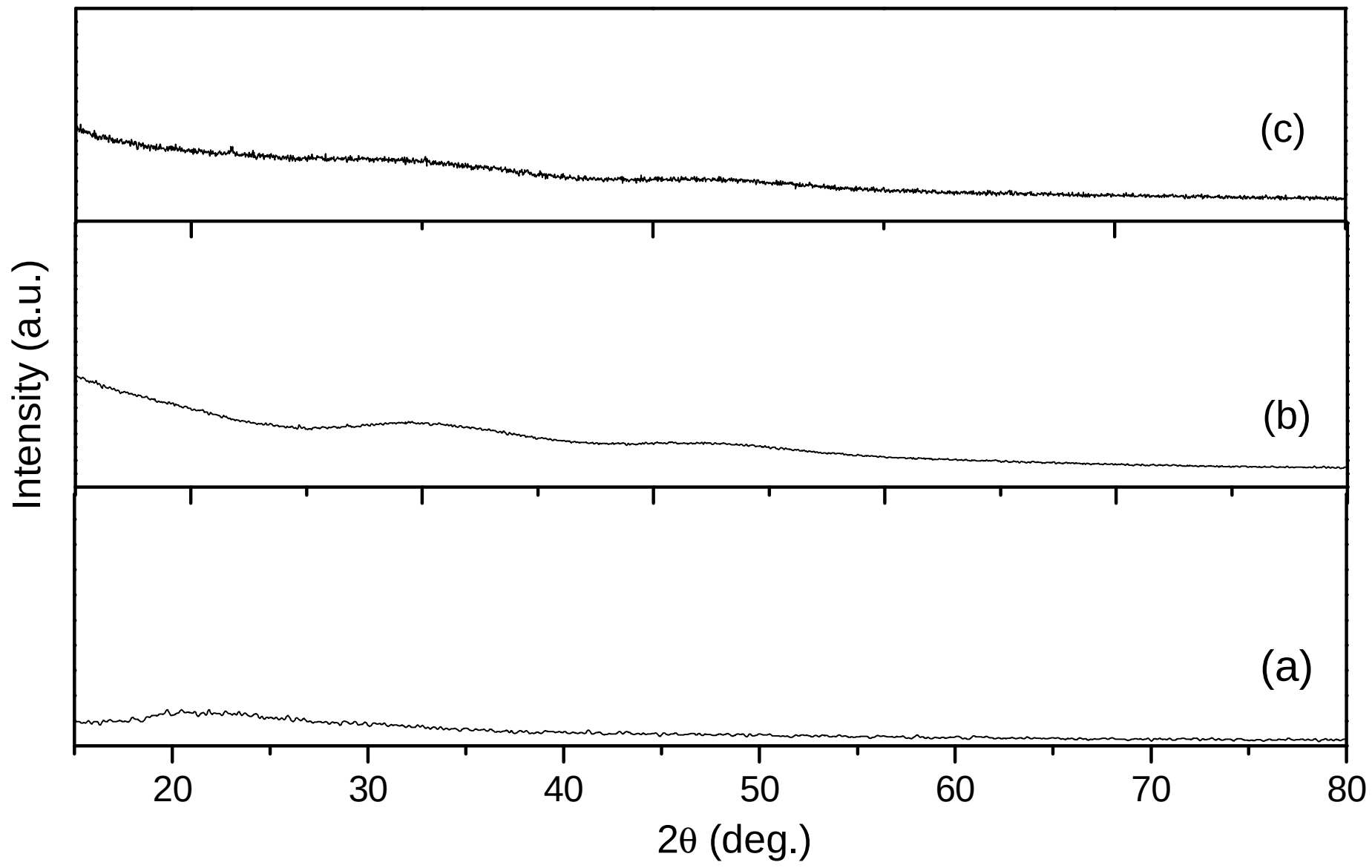

Fig.1: X-ray powder diffraction patterns for the as-quenched glasses of the (a) $Li_2O$-$2B_2O_3$, (b) $0.5Li_2O$-$0.5Na_2O$-$2B_2O_3$, (c) $0.5Li_2O$-$0.5K_2O$-$2B_2O_3$

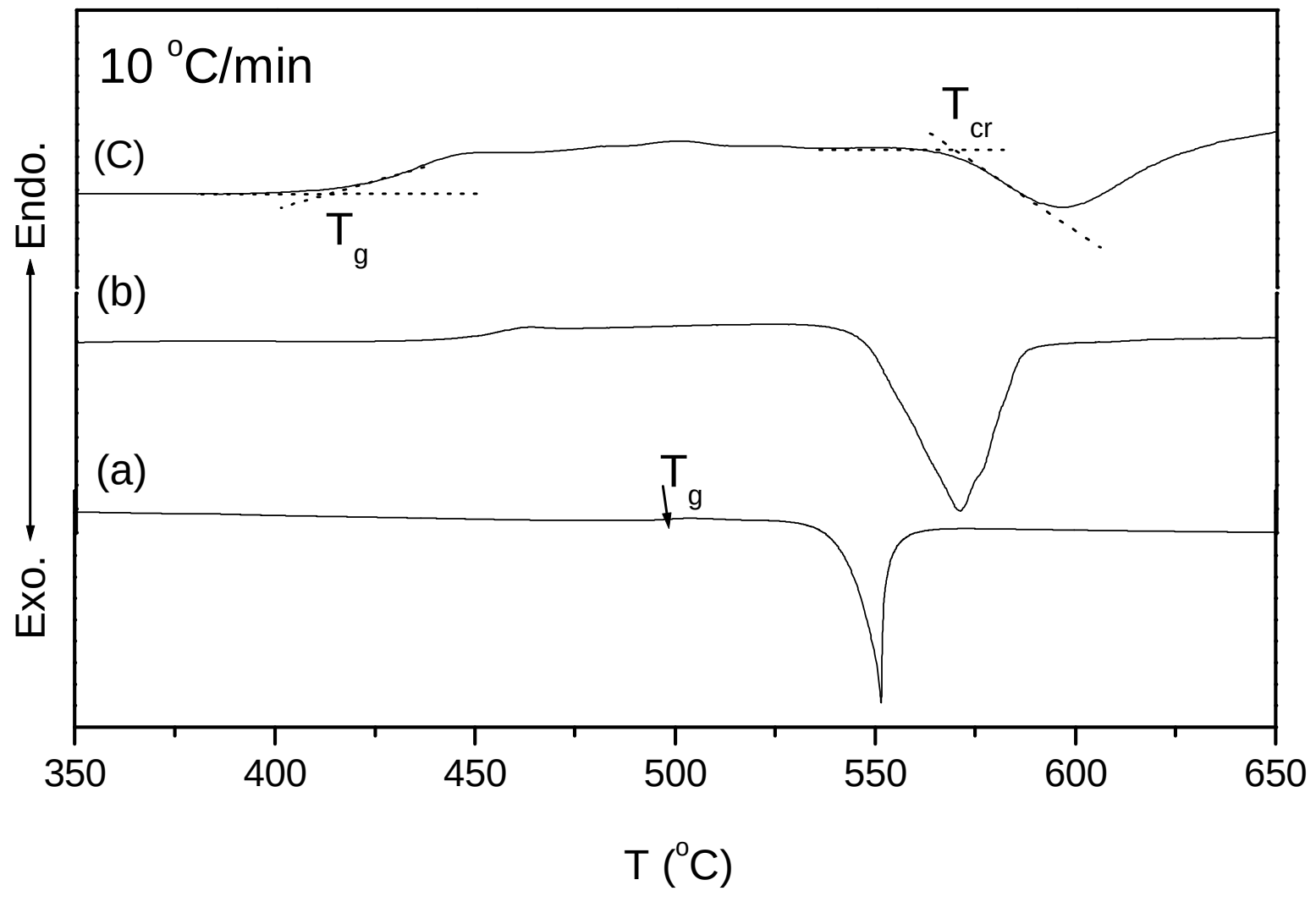

Fig.2: DSC traces for the as-quenched glasses of the (a) $Li_2O$-$2B_2O_3$, (b) $0.5Li_2O$-$0.5Na_2O$-$2B_2O_3$, (c) $0.5Li_2O$-$0.5K_2O$-$2B_2O_3$

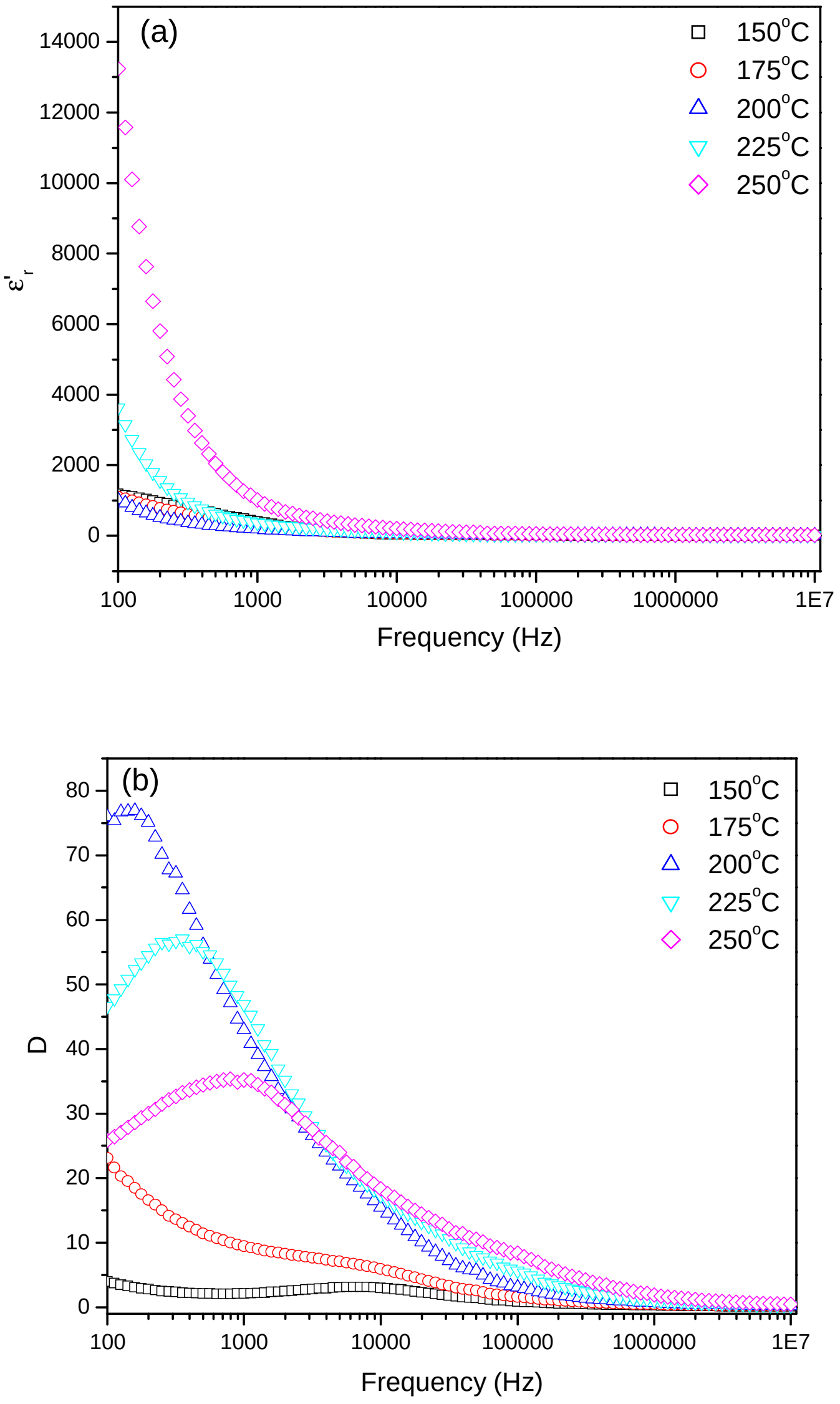

Fig.3 Frequency dependent (a) dielectric constant
(b) dielectric loss at various temperatures

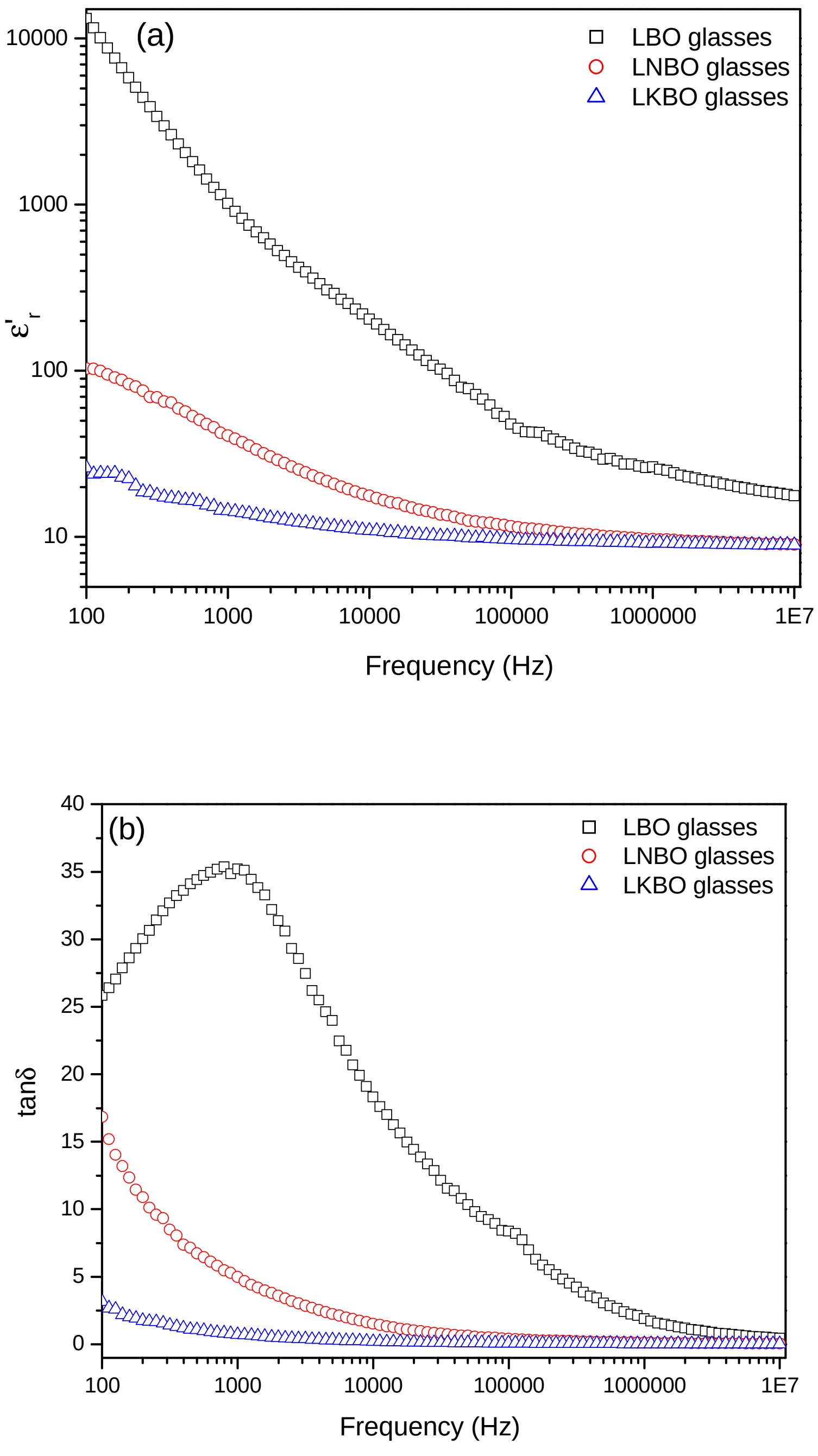

Fig. 4: (a) Frequency dependent dielectric constant and (b) loss at 250$^{o}$C for the three different glasses

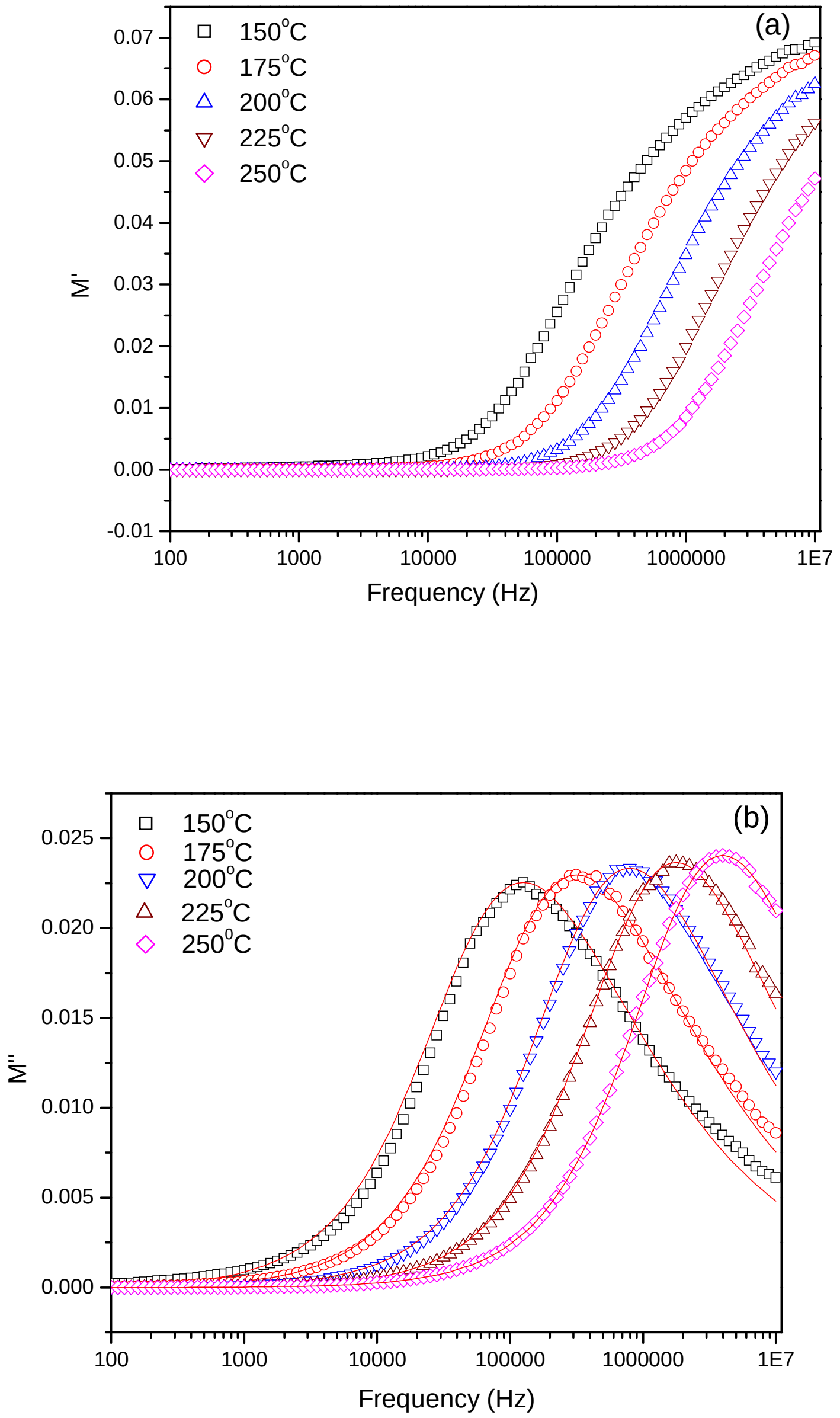

Fig.5: (a) Frequency dependent real and (b) imaginary parts of electric modulus (solid lines are theoretical fitted curves) for LBO glasses

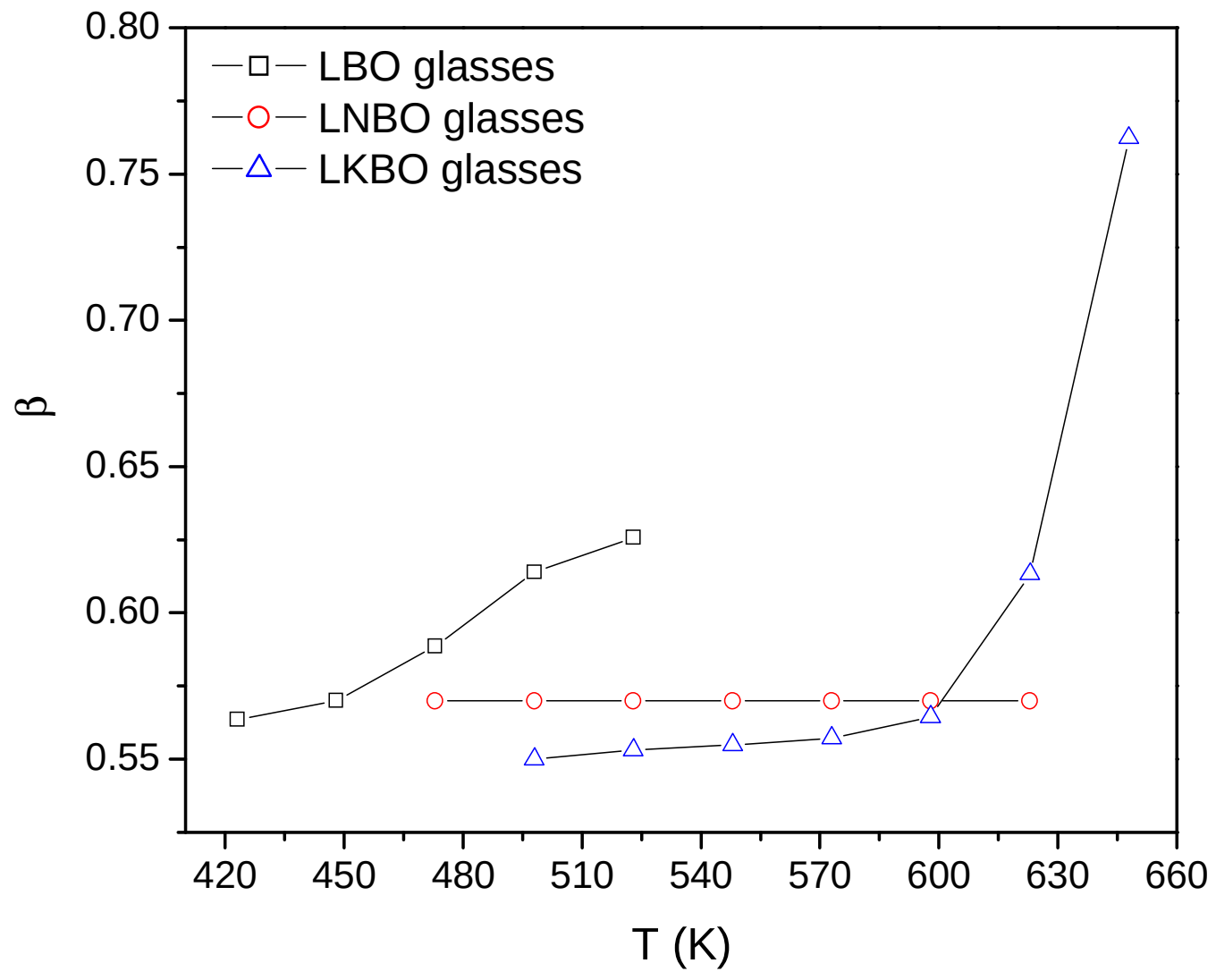

Fig.6: β vs temperature plots for LBO, LNBO and LKBO glasses

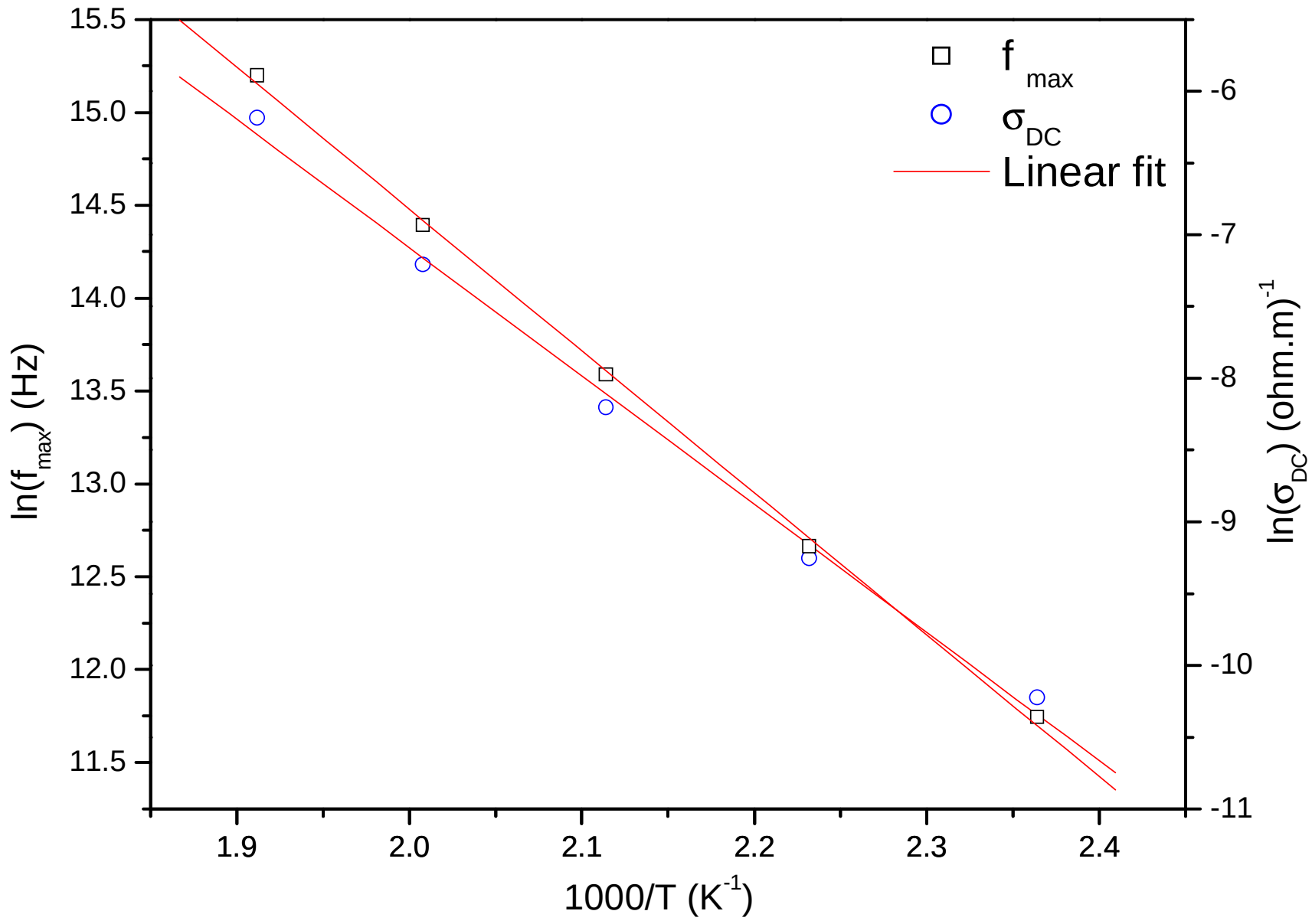

Fig.7: Arhenious plots of DC conductivity and relaxation frequency for LBO glasses

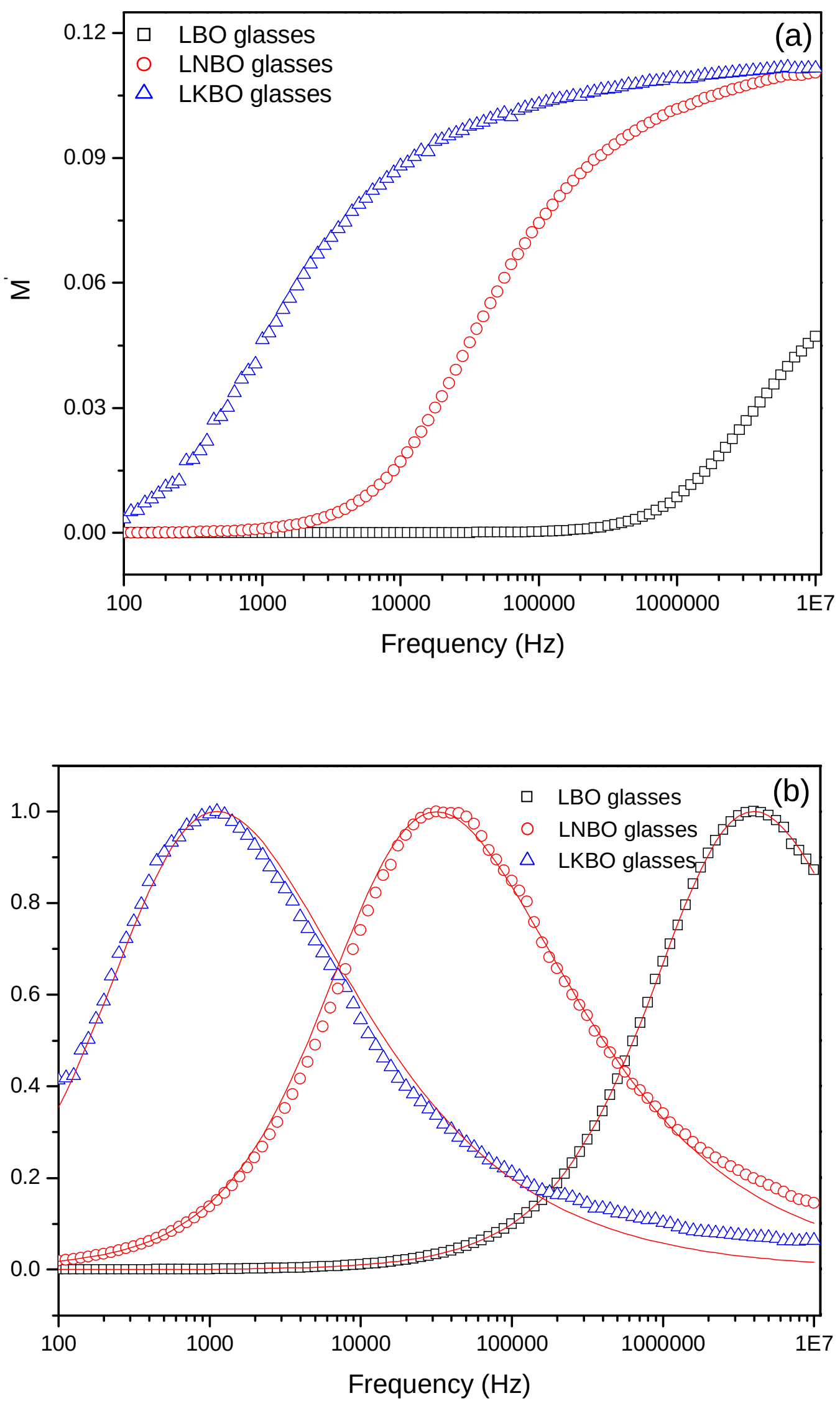

Fig.8: (a) Frequency dependent real and (b) imaginary parts of electric modulus at $250^0$C (solid lines are theoretical fitted curves) for LBO, LNBO and LKBO glasses

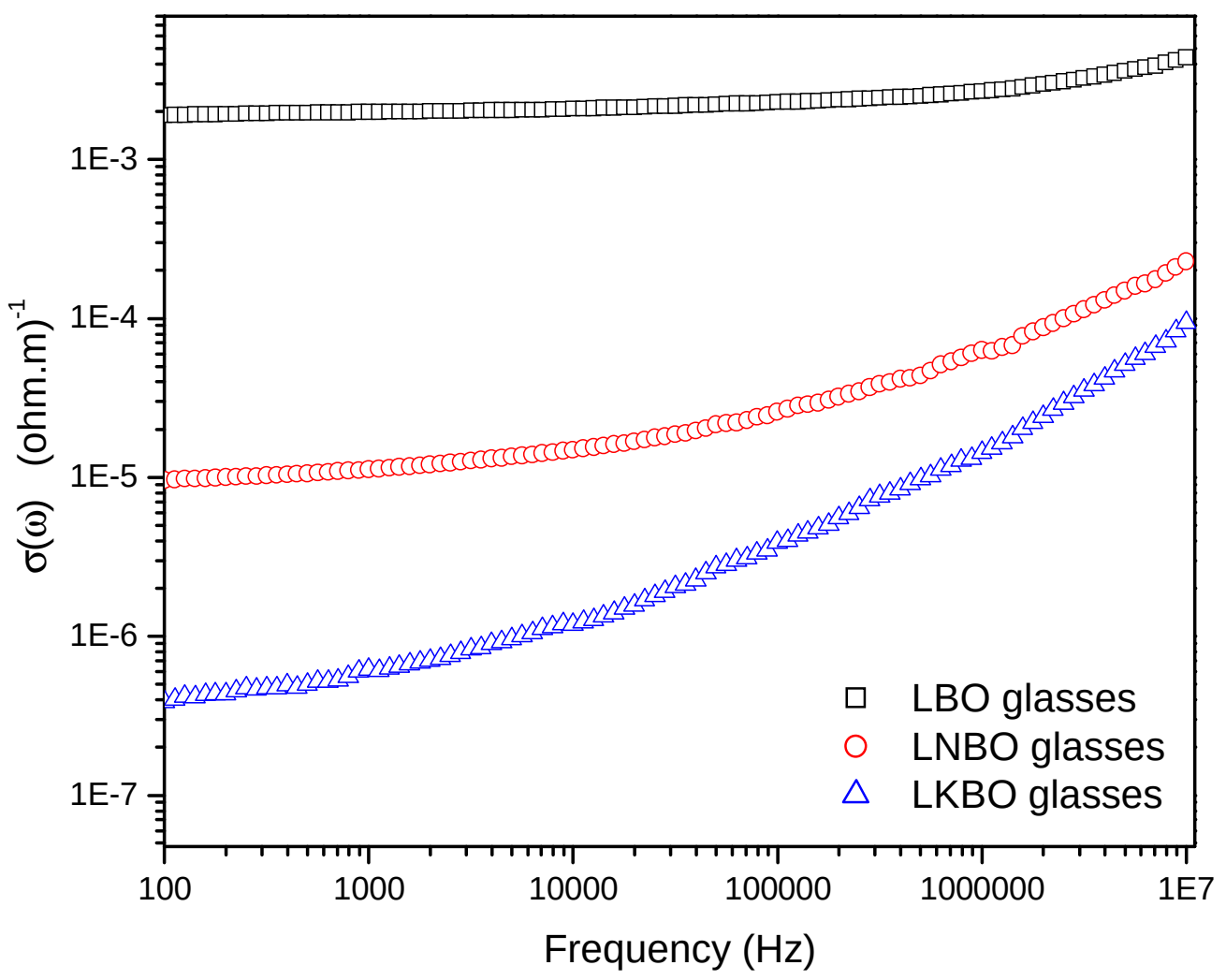

Fig. 9: Frequency dependent ac conductivity at 250°C for LBO, LNBO and LKBO glasses